\documentclass[aoas,nameyear,dvips]{arximspdf}
\usepackage{dcolumn}
\usepackage{graphicx}

\doi{10.1214/09-AOAS278}
\volume{4}
\issue{1}
\pubyear{2010}
\firstpage{439}
\lastpage{459}

\makeatletter
\newcolumntype{d}[1]{D{.}{.}{#1}}
\makeatother

\begin{document}
\begin{frontmatter}

\title{A latent factor model for spatial data\\ with informative
missingness\thanksref{TZ}}
\runtitle{A model for spatial data with informative missingness}
\thankstext{TZ}{Supported in part by NIH/NCRR Grant P20 RR017696-06 and
NIH Grant 1R01LM009153.}

\begin{aug}
\author[A]{\fnms{Brian J.} \snm{Reich}\corref{}\ead[label=e1]{reich@stat.ncsu.edu}}\and
\author[B]{\fnms{Dipankar} \snm{Bandyopadhyay}\ead[label=e2]{bandyopd@musc.edu}}
\runauthor{B. J. Reich and D. Bandyopadhyay}
\affiliation{North Carolina State University and Medical University of
South Carolina}
\address[A]{Department of Statistics\\
North Carolina State University\\
2311 Stinson Drive\\
4264 SAS Hall, Box 8203\\
Raleigh, North Carolina 27695\\
USA\\
\printead{e1}}
\address[B]{Center for Oral Health Research\\
Division of Biostatistics and  Epidemiology\\
Department of Medicine\\
Medical University of South Carolina\\
135 Cannon Street, Suite 303\\
Charleston, South Carolina 29425\\
USA\\
\printead{e2}}
\end{aug}

\received{\smonth{5} \syear{2009}}
\revised{\smonth{7} \syear{2009}}

%
\begin{abstract}
A large amount of data is typically collected during a periodontal exam.
Analyzing these data poses several challenges. Several types of measurements
are taken at many locations throughout the mouth. These spatially-referenced
data are a mix of binary and continuous responses, making joint
modeling difficult.
Also, most patients have missing teeth. Periodontal disease is a
leading cause of
tooth loss, so it is likely that the number and location of missing
teeth informs
about the patient's periodontal health. In this paper we develop a multivariate
spatial framework for these data which jointly models the binary and continuous
responses as a function of a single latent spatial process representing
general periodontal health.
We also use the latent spatial process to model the location of missing teeth.
We show using simulated and real data that exploiting spatial
associations and jointly
modeling the responses and locations of missing teeth mitigates the
problems presented
by these data.
\end{abstract}

%
\end{frontmatter}

\section{Introduction}\label{s:intro}

\begin{figure}

\includegraphics{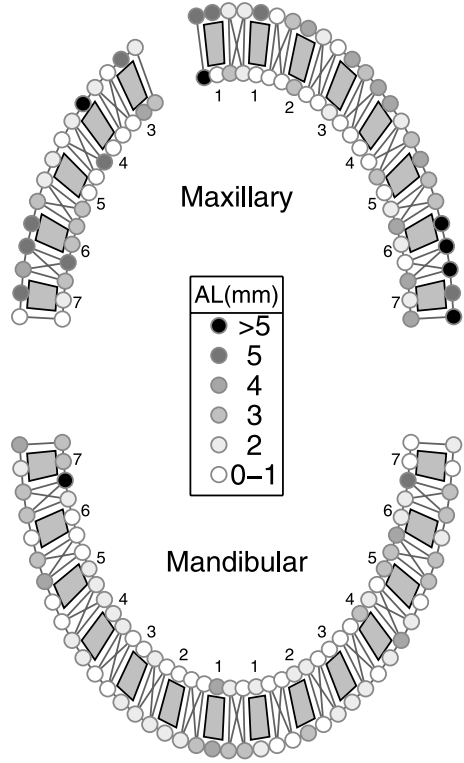}

\caption{Observed CAL for a typical patient. The shaded boxes represent
teeth, the circles represent measurement sites, and the gray lines represent
neighbor pairs connecting adjacent sites on the same tooth and sites that
share a gap between teeth. ``Maxillary'' and ``Mandibular'' refer to
upper and
lower jaws respectively. The small numbers beside each tooth are the ``tooth
numbers.'' The maxilla's second tooth on the left is missing; third molars
(``wisdom teeth'') are excluded.}\label{f:data}
\end{figure}

Periodontal disease or periodontitis is an inflammatory disease
affecting periodontium, the tissues that support and maintain teeth.
Periodontitis causes progressive bone loss around the tooth which can
lead to tooth loosening and eventually tooth loss. It has been
estimated that about 50\% of US adults over the age of 35 experience
early stages of periodontal disease [Oliver, Brown and Loe (\citeyear
{OliverBrownLoe1998})],
making periodontitis the primary cause of adult tooth loss. To measure
periodontal status, dental hygienists often use a periodontal probe to
measure several disease markers throughout the mouth. Three of the most
popular markers are (a) clinical attachment loss (CAL), (b) periodontal
pocket depth (PPD) and (c) bleeding on probing (BOP). PPD and CAL are
continuous variables, usually rounded to the nearest millimeter. CAL is
the distance down a tooth's root that is no longer attached to the
surrounding bone by the periodontal ligament, and PPD is the distance
from the gingival margin to the base of the pocket. BOP is a binary
response and is indicative of whether a particular site bled with the
application of a dental probe. During a full periodontal exam, all
three markers are usually measured at six pre-specified sites
[Darby and Walsh (\citeyear{DarbyWalsh1995})] for each tooth
(excluding the third molars, i.e., the
wisdom teeth). So for a patient with no missing teeth, there are
$S=168$ measurements for each marker (Figure~\ref{f:data}).

%

The motivating example is a clinical study conducted at the Medical
University of South Carolina (MUSC) to determine the periodontal
disease status for Type-2 diabetic Gullah-speaking African-Americans,
originally presented in Fernandez et al. (\citeyear
{FernandezEtAl2009}). The objective of this
analysis is to quantify the disease status of this population, and to
study the associations between disease status and patient-level
covariates such as age, BMI, gender, HbA1C and smoking status.

Quantifying a patient's disease status from the extensive data
collected during a periodontal exam is difficult. For example, it is
common to summarize disease status using the whole-mouth average CAL or
the number of teeth with CAL above a certain threshold. Using the
whole-mouth average CAL as the response in a regression with
patient-level covariates is reasonable when the patients' residual
distributions are identical. However, this assumption is often violated
in practice, as different patients have different error variances,
spatial covariances and missing data patterns. In this paper we present
a multivariate spatial model to jointly analyze periodontal data from
multiple markers and multiple measurement locations to improve
estimation of disease status, and hence develop a more powerful method
for studying the association between patient-level covariates and
periodontal disease.

We use spatial factor analysis [Wang and Wall (\citeyear{WangWall2003}),
Hogan and Tchernis (\citeyear{HoganTchernis2004}),
Lopes, Salazar and Gamerman (\citeyear{LopesSalazarGamerman2008})] to
model these multivariate
spatial data. We postulate that the three markers are all related to a
single latent spatial process (factor) measuring periodontal health.
The latent periodontal health factor varies from site to site and is
smoothed spatially using a conditionally autoregressive prior
[Besag, York and Molli\'{e} (\citeyear{BesagYorkMolli991}),
Banerjee, Carlin and Gelfand (\citeyear{BanerjeeCarlinGelfand2004})].
The data collected for this study
provide interesting challenges that require extensions of the spatial
factor model. First, the data are a mix of continuous and binary
responses. To jointly model these data types, we develop a spatial
probit model for binary responses, which has the advantage of being
fully-conjugate and leads to rapid MCMC sampling and convergence. Also,
we have data from multiple patients, and exploratory analysis suggests
that the covariance of the latent spatial factor varies by patient.
Therefore, we develop a hierarchial model which allows the covariance
to vary between patients, but pools information across patients to
estimate the covariance parameters. We show in a simple case that in
terms of estimating the effect of patient-level covariates, this model
is equivalent to a weighted multiple regression, where the patient's
scalar response is a linear combination of all data across location and
marker, and the patient's weight decreases with the spatial correlation
and variability.

Another challenging aspect of analyzing periodontal data is the
considerable number of missing teeth (around 20\% for these data). The
assumption that teeth are missing completely at random is not valid
because periodontal disease is the leading cause of adult tooth loss,
so patients with several missing teeth likely have poor periodontal
health. For nonspatial data a common method to handle so-called
``informative cluster size'' is to include the number of observations
as a covariate, or in the weights of a weighted regression
[Hoffman, Sen and Weinberg (\citeyear{HoffmanSenWeinberg2001}),
Williamson, Datta and Satten (\citeyear{WilliamsonDattaSatten2003}),
Follman, Proschan and Leifer (\citeyear{FollmanProschanLeifer2003}),
Lu (\citeyear{Lu2005}),
Benhin, Rao and Scott (\citeyear{BenhinRaoScott2005}),
Panageas et al. (\citeyear{PanageasEtAl2007}),
Cong, Yin and Shen (\citeyear{CongYinShen2007}),
Williamson et al. (\citeyear{WilliamsonEtAl2008})].
Dunson, Chen and Harry (\citeyear{DunsonChenHarry2003}) take a
different approach. They propose a joint
model for clustered mixed (continuous and binary) data and the number
of responses in each cluster, using a continuation ratio probit model
for cluster size. Another approach is the shared parameter model
[e.g., Wu and Carroll (\citeyear{WuCarroll1988}),
Follman and Wu (\citeyear{FollmanWu1995})]. The shared parameter
model accounts for informative missingness by introducing random
effects that are shared between the missing data process and the
measurement process. Conditioned on the random effects, the missing
data and measurement processes are assumed to be independent.

We propose a shared variable model to jointly model missing teeth with
the other markers of periodontal disease. However, in our spatial
setting both the number \textit{and} location of missing teeth are
informative. For example, a missing tooth in the front of the mouth
surrounded by teeth with low CAL may not be informative; in contrast, a
missing tooth in the back of the mouth (where periodontal disease is
often the most advanced) surrounded by teeth with high CAL is
indicative of poor periodontal health in that region of the mouth.
Therefore, we model the number and spatial distribution of missing
teeth using our latent spatial factor model. In this model, CAL, PPD,
BOP and the location of missing teeth are all modeled simultaneously in
terms of a latent periodontal health factor; this approach uses all
available information to estimate periodontal disease status.

The paper proceeds as follows. Section~\ref{s:model} presents our
unified approach to modeling multivariate spatially-referenced
periodontal data, as well as our model for informatively missing teeth.
Section~\ref{s:diag} offers some influence diagnostics to determine
which patients and response types are the most informative about the
patient-level covariates. Computing details are given in Section~\ref
{s:mcmc}. Section~\ref{s:sim}'s simulation study shows that accounting
for spatial association and informative observation location can lead
to a substantial improvement in estimating the patient-level covariate
effects. We analyze the periodontal data in Section~\ref{s:results}.
Section~\ref{s:disc} concludes.

\section{Latent spatial factor model for periodontal data}\label{s:model}
In this section we describe our approach for spatially-referenced mixed
periodontal data with informative missingness. We begin in Section~\ref
{s:complete} by specifying a latent spatial factor model assuming no
missing teeth. Section~\ref{s:missing} introduces the spatial probit
model for missing teeth and Section~\ref{s:ID} specifies priors and
discusses identifiability of the latent variable model.

\subsection{Complete data model}\label{s:complete}
We assume our multivariate spatial data has $J$ types of responses (for
the periodontal data the $J=3$ responses are CAL, PPD and BOP) at each
spatial location for each patient. If the $j$th type of response is
continuous (CAL and PPD), let $y_{ij}(s)$ be the response at spatial
location $s$ for patient $i$, $s=1,\ldots,S$ and $i=1,\ldots,N$. Our
data also has binary responses (BOP). If the $j$th type of response
is binary, let $y^*_{ij}(s)$ be the response at spatial location $s$
for patient $i$. We model binary responses using probit regression,
that is, $y^*_{ij}(s) = I(y_{ij}(s)>0)$, where $I(\cdot)$\vspace*{1pt} is the binary
indicator function and $y_{ij}(s)$ is a Gaussian latent variable.

All $J$ responses are modeled as functions of the latent spatial
disease status, $\mu_{i}(s)$, which represents the overall periodontal
health of patient $i$ at location $s$. Let
%
\begin{equation}\label{ymodel1}
y_{ij}(s) = a_{j} + b_{j}\mu_{i}(s) + \varepsilon_{ij}(s),
\end{equation}
where $a_{j}$ is the intercept for response $j$, $b_{j}$ relates the
latent factor to response type~$j$, and $\varepsilon_{ij}(s)\sim$
$N(0,\sigma_{ij}^2$) is error. As is customary for probit regression,
we assume $\sigma_{ij}^2=1$ for binary responses for identification.
Since all $J$ responses depend on the common latent factor, they are
correlated with
%
\begin{equation}\label{corry}
\operatorname{Cor} (y_{ij}(s),y_{il}(s) ) =
\frac{b_jb_l\operatorname{Var}[\mu_{i}(s)]}
{\sqrt{b_j^2\operatorname{Var}[\mu_{i}(s)]+\sigma^2_{ij}}\sqrt
{b_l^2\operatorname{Var}[\mu
_{i}(s)]+\sigma^2_{il}}}.
\end{equation}
The slopes $b_j$ and $b_l$ determine the sign and magnitude of the
correlation; if either $b_j$ or $b_l$ is zero, then $y_{ij}(s)$ and
$y_{il}(s)$ are uncorrelated, and if $b_j$ and $b_l$ share (do not
share) the same sign, then $y_{ij}(s)$ and $y_{il}(s)$ are positively
(negatively) correlated.

The latent vector $\bolds{\mu}_{i} = (\mu_i(1),\ldots,\mu_i(S))'$
has a
multivariate normal prior with conditionally autoregressive covariance
[``CAR,'' Besag, York and Molli\'{e} (\citeyear{BesagYorkMolli991})].
The mean of $\bolds{\mu}_{i}$ is
\[
E(\bolds{\mu}_{i})=W\bolds{\alpha}+\Omega_i\bolds{\beta},
\]
where $W$ is an $S \times q$ matrix of spatial covariates (e.g., tooth
number) that do not vary across patient, $\mathbf{X}_i$ is the
$p$-vector of
patient-level covariates (e.g., age) that do not vary across space
within patient, $\Omega_i=X_i'\otimes\mathbf{1}_S$, $\mathbf{1}_S$ is
the $S$-vector of ones, and $\bolds{\alpha}$ and $\bolds{\beta}$
are the
corresponding regression parameters. The covariance of $\bolds{\mu
}_i$ is
$\tau
_i^2Q(\rho_i)^{-1}$, where $Q(\rho_i) = M-\rho_iD$, $D_{ss'}$ is one if
locations $s$ and $s'$ are adjacent and zero otherwise, $M$ is diagonal
with diagonal elements $M_{ss}=\sum_{s'}D_{ss'}$. In this spatial
model, $\rho_i\in[0,1]$ controls the degree of spatial association and
$\tau_i^2 >0$ controls the magnitude of variation. Let $r_i(s) = \mu
_{i}(s)-E(\mu_{i}(s))$. A convenient interpretation of the CAR prior is
that the conditional distribution of $r_i(s)$ given $r_i(s')$ for all
$s'\ne s$ is normal with mean $\rho_i\bar{r}_i(s)$ and variance $\tau
_i^2/m(s)$, where $\bar{r}_i(s)$ is the average of $r_i(s)$ at location
$s$'s $m(s)$ neighbors.

The degree of spatial variation is allowed to differ between patients
by means of $\sigma_{ij}^2$, $\tau_i^2$ and $\rho_i$. To pool
information across patients, we use models
%
\begin{eqnarray}\label{priors}
\sigma_{ij}^{-2}| c_j, d_j & \sim& \operatorname{Gamma}(c_j,d_j),\nonumber\\
\tau_{i}^{-2}| e, f & \sim& \operatorname{Gamma}(e,f),\\
\rho_{i}| g, h & \sim& \operatorname{Beta}(g,h),\nonumber
\end{eqnarray}
where $\{c_j\}$, $\{d_j\}$, $e$, $f$, $g$ and $h$ are hyperparameters.

\subsection{Model for the location of missing teeth}\label{s:missing}
For our data described in Section~\ref{s:intro}, roughly 20\% of the
teeth are missing. The locations of the missing teeth are not random,
but rather related to the periodontal health in that region of the
mouth. Therefore, we propose a model for the location of missing teeth
as a function of the underlying latent factor $\mu_i(s)$.

For our data either the six observations on a tooth for all $J$
responses are all observed or all unobserved. That is, if a tooth is
missing, we have no data for the tooth, and if a tooth is not missing,
we have complete data. Let $y^*_{i0}(t)$ be an indicator of whether
tooth $t=1,\ldots,T$ is missing for patient $i$. As with the binary
data in Section~\ref{s:model}, we model $y^*_{i0}(t)$ using probit
regression. Let $y^*_{i0}(t)=I(y_{i0}(t)>0)$, where $y_{i0}(t)$ is a
latent continuous variable. Then
%
\begin{equation}
y_{i0}(t) = a_{0}+b_{0}Z_t'\bolds{\mu}_i + \varepsilon_{i0}(t),
\end{equation}
where $Z_t$ is such that $Z_t'\bolds{\mu}_i$ is the mean of $\bolds{\mu}_i$ at the
six observations on tooth $t$ and $\varepsilon_{i0}(t)\stackrel{\mathrm{i.i.d.}}{\sim}N(0,1)$. $a_{0}$ and $b_{0}$
relate the latent process to
the missing tooth indicator. Note that since $\mu_i(s)$ is included in
both the model for presence of and value of the responses, both
presence and value of the data contribute to the posterior of $\mu_i(s)$,
and thus the posterior of $\bolds{\beta}$. Also note that $b_{i0}=0$
corresponds to independence between the latent factor and the location
of missing teeth, in which case the location of missing teeth does not
contribute to estimating $\bolds{\beta}$.

\subsection{Identifiability and prior choice}\label{s:ID}

Identifiability is a key issue in latent variable modeling. To see
this, we inspect the first two moments of the multivariate response at
location $s$ for patient $i$ after integrating over the latent factor
$\bolds{\mu}_i$,
%
\begin{eqnarray}\label{ID}
\mbox{E} (y_{ij}(s) ) &=& a_j + b_j [W(s)\bolds{\alpha}+\mathbf{X}_i'\bolds{\beta}],\nonumber\\[-8pt]\\[-8pt]
\operatorname{Cov} (y_{ij}(s),y_{il}(s) ) &=& b_jb_l\tau_i^2q(s) +I(j=l)\sigma^2_{ij},\nonumber
\end{eqnarray}
where $W(s)$ is the row of $W$ corresponding to location $s$ and $q(s)$
is the $(s,s)$ diagonal element of $Q(\rho_i)^{-1}$. Identifiability
concerns arise in both moment expressions, as multiplying all of the
slopes $b_j$ by scalar $c$ and dividing $\bolds{\alpha}$, $\bolds
{\beta}$ and $\tau
_i^2$ by $c$ gives identical moments. Although there are other ways to
address this issue, we fix $b_1\equiv1$. This identifies both the
regression coefficients $\bolds{\alpha}$ and $\bolds{\beta}$ via
the mean of the
first response and the CAR variance $\tau_i^2$ via the variance of the
first response. In our analysis of periodontal data of Section~\ref
{s:results} we take the first response with fixed slope to be clinical
attachment loss, the most commonly used measure of periodontal disease.
We also compare these results with other baseline assignments and
discuss sensitivity to this assumption.

The regression coefficients $\{a_j\}$, $\{b_j\}$ ($j\ne1$), $\bolds
{\alpha}$
and $\bolds{\beta}$ have independent $N(0, w^2)$ priors. The hyperparameters
$\{c_j\}$, $\{d_j\}$, $e$, $f$, $g$ and $h$ have independent
Gamma($u,v$) priors. In the simulation study (Section~\ref{s:sim}) and
data analysis (Section~\ref{s:results}) we take $u=v=0.1$ and $w=10$ to
give vague yet proper priors. We conduct a sensitivity analysis in
Section~\ref{s:results} which shows that the results are not sensitive
to these priors for this large periodontal data set.

\section{Influence diagnostics}\label{s:diag}
Our primary interest is in the patient-level parameters $\bolds{\beta
}$. In
this complicated hierarchical Bayesian model, we would like to identify
the sources of data that are most informative about $\bolds{\beta}$.
In this
section we develop diagnostics to determine which patients, spatial
locations and response types are the most influential. We assume no
missing teeth, that all responses are Gaussian, and that no covariates
depend on space ($W$ is null). In this case the regression coefficients
only affect the overall average response for each patient, and a
tempting simplification is to collapse data over space and use each
patient's overall average as a scalar response. We show that even in
this case different areas of the mouth are more than less informative,
and that patients are weighted differently depending on their spatial
covariance parameters. This motivates the hierarchical spatial model
even in this simple case.

Integrating over latent effect $\bolds{\mu}_i$, but conditioning on
$\sigma
^2_{ij}$, $\tau^2_i$, $\rho_{i}$, $a_{j}$ and $b_{j}$, the posterior
for $\bolds{\beta}$ is Gaussian with
%
\begin{eqnarray}\label{weightedreg}
\operatorname{COV}(\bolds{\beta}) &=& \Biggl(\sum_{i=1}^N w_i\mathbf{x}_i\mathbf{x}_i'\Biggr)^{-1},\nonumber\\[-8pt]\\[-8pt]
E(\bolds{\beta}) &=& \Biggl(\sum_{i=1}^N w_i\mathbf{x}_i\mathbf{x}_i'\Biggr)^{-1}\sum_{i=1}^N w_i\mathbf{x}_i'z_i,\nonumber
\end{eqnarray}
where
%
\begin{equation}\label{w}
w_i=\tau_i^{-2}\mathbf{1}' \bigl[Q(\rho_i)-Q(\rho_i)\bigl(\delta_iI_S+Q(\rho
_i)\bigr)^{-1}Q(\rho_i) \bigr]\mathbf{1},
\end{equation}
$\delta_i = \tau_i^2\sum_{j=1}^Jb_{j}^2/\sigma_{ij}^2$, and
\[
z_i=\frac{1}{w_i}\mathbf{1}'Q(\rho_i)\bigl(\delta_iI_s+Q(\rho_i)\bigr)^{-1}
\sum_{j=1}^Jb_{j}\sigma_{ij}^{-2}(\mathbf{y}_{ij}-a_{j}).
\]
The posterior in (\ref{weightedreg}) is equivalent to a weighted linear
regression where each patient contributes the scalar response $z_i$ and
is weighted according to $w_i$. Analyzing $z_i$ and $w_i$ shows which
sites, patients and outcomes contribute the most to $\bolds{\beta}$'s
posterior.

\begin{figure}

\includegraphics{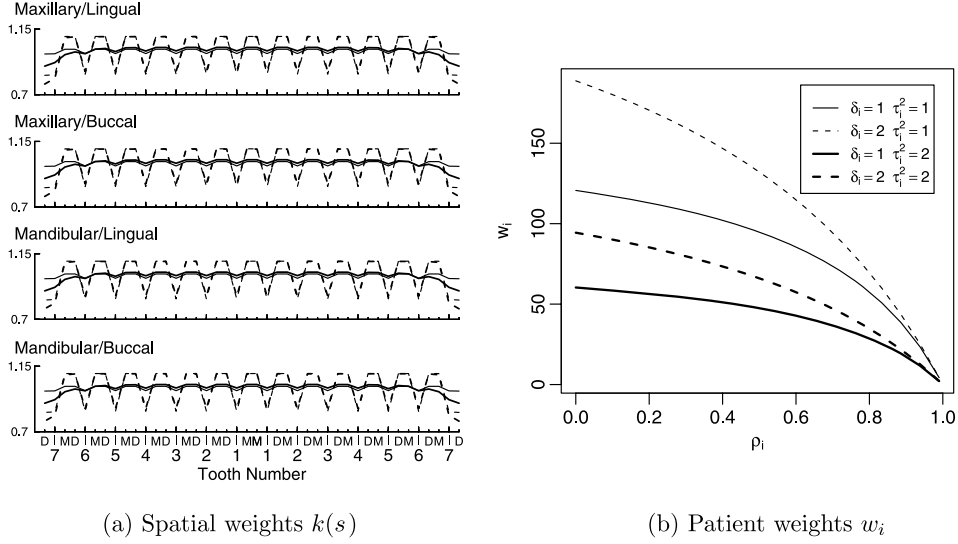}

\caption{Panel \textup{(a)} plots the spatial weights $k(s)$ for various
$\delta$ and $\rho$. ``Maxillary'' and ``Mandibular'' refer to upper and lower
jaws respectively, while ``buccal'' and ``lingual'' refer to the cheek
and the tongue sides of the teeth, respectively. The thin lines have
$\rho=0.1$, the wide lines have $\rho=0.99$; the solid lines have
$\delta=0.2$, the dashed lines have $\delta=5$. Panel \textup{(b)} plots the
patient weights $w_i$ for various $\delta_i$, $\tau_i$ and $\rho_i$.}\label{f:diags}
\end{figure}

First we consider $z_i$:
%
\begin{eqnarray}\label{z}
z_i&=&\frac{1}{w_i}
\mathbf{1}' \bigl[Q(\rho_i)\bigl(\delta_iI_s+Q(\rho_i)\bigr)^{-1} \bigr]
\sum_{j=1}^J\frac{b_j}{\sigma_{ij}^2}(\mathbf{y}_{ij}-a_j)\nonumber\\[-8pt]\\[-8pt]
&=&
\sum_{j=1}^J\frac{b_j}{\sigma_{ij}^{2}} \Biggl[\sum_{s=1}^Sk_i(s)\bigl(y_{ij}(s)-a_j\bigr) \Biggr],\nonumber
\end{eqnarray}
where the vector $k_i=\mathbf{1}' [Q(\rho_i)(\delta_iI_s+Q(\rho
_i))^{-1} ]/w_i$. Therefore, $z_i$ is a linear combination of all the
observations for patient $i$, with $k(s)$ controlling the relative
weight of observations at location $s$ and $b_j/\sigma_{ij}^{2}$
controlling the relative weight of response type $j$. Figure~\ref
{f:diags}(a) plots $k(s)$ (scaled to sum to $S$) for four combinations of
$\rho_i$ and $\delta_i$. Observations in the gaps between teeth have
the highest weight; these sites have the most neighbors and thus the
smallest prior variance. Observations in the back of the mouth and on
the sides of teeth get less weight.

%

The patient weights $w_i$ are plotted as a function of $\rho_i$,
$\delta
_i$ and $\tau_i$ in Figure~\ref{f:diags}(b). The weight decreases with
$\rho_i$ and $\tau_i^2$, and increases with $\delta_i$ (inversely
related to error variances $\sigma_{ij}^2$). That is, patients with
little spatial association and small variances $\tau_i^2$ and
$\sigma_{ij}^2$\vspace*{1pt} (and thus large $\delta_i$) have the most influence on
$\bolds{\beta}
$'s posterior.
To search for overly-influential patients, we compute the weights by
evaluating (\ref{w}) using posterior means ${\hat\rho}_i$, ${\hat
\tau
}_i^2$ and ${\hat\sigma}_{i1}^2$. However, the marginal posterior for
$\bolds{\beta}$ is not available in closed-form for Section~\ref{s:results}'s
data with informative missing teeth and binary responses. Therefore, we
use only the CAL error variance ${\hat\sigma}_{i1}^2$, that is,
$\delta
_i={\hat\tau}_i^2/{\hat\sigma}_{i1}^2$ ($b_1=1$, Section~\ref{s:ID}),
as an approximation. This approximation is not meant to be definitive,
but rather a useful heuristic device.

\section{MCMC sampling algorithm}\label{s:mcmc}

MCMC sampling is carried out using the free software \textit{R}
(\url{http://www.r-project.org/}), although it would also be straightforward
to implement the model using WinBUGS\break
(\url{http://www.mrc-bsu.cam.ac.uk/bugs/}). Sample code to analyze a
single continuous response is available in the supplemental article
[Reich and Bandyopadhyay (\citeyear{ReichBandyopadhyay2009})]. We draw
20,000 MCMC samples and
discard the first 5000 as burn-in. Convergence is monitored using
trace plots of the deviance as well as several representative parameters.

The patient-specific parameters are conditionally-conjugate except for
the CAR spatial association parameters $\rho_i$, which are updated
using Metropolis--Hastings sampling with a Beta$ (50\rho_i^*, 50(1-\rho_i^*) )$
candidate distribution, where $\rho_i^*$ is the value at the
previous iteration. The remaining parameters are updated using Gibbs
sampling with full conditionals given below. The latent continuous
variables corresponding to the probit model for binary responses,
$y_{ij}(s)$, are updated from their truncated full conditionals
$y_{ij}(s) \sim N(a_{j}+b_{j}\mu_{i}(s),1)$, restricted to $(-\infty
,0)$ if $y^*_{ij}(s)=0$ and $(0,\infty)$ if $y^*_{ij}(s)=1$. The vector
of latent effects for patient $i$, $\bolds{\mu}_i$, is multivariate
normal with\vspace*{-2pt}
\begin{eqnarray*}
V(\bolds{\mu}_i|\operatorname{rest})^{-1} &=& \mathbf{Z}'\mathbf{Z}b_{0}^2+Q(\rho_i)/\tau_i^2
+\Biggl(\sum_{j=1}^Jb_{j}^2/\sigma_{ij}^2 \Biggr)I_n,\nonumber\\
E(\bolds{\mu}_i|\operatorname{rest}) & = & V(\bolds{\mu}_i|\operatorname{rest})\Biggl(b_{0}\mathbf{Z}'(\mathbf{y}_{0i}-a_{0})
\\
&&\hspace*{54pt}{}+Q(\rho_i)(W\bolds{\alpha}+\Omega_i\bolds{\beta})/\tau_i^2+\sum_{j=1}^J
b_{j}(\mathbf{y}_{ij}-a_j)/\sigma_{ij}^2 \Biggr),\nonumber\vspace*{-1pt}
\end{eqnarray*}
where $\mathbf{Z}=(Z_1,\ldots,Z_T)$, $\mathbf{y}_{ij} =(y_{ij}(1),\ldots,y_{ij}(S))$
and $\mathbf{y}_{i0} = (y_{i0}(1),\ldots,\break y_{i0}(T) )$. The measurement
error variances for the continuous responses have full conditional\vspace*{-2pt}
%
\begin{eqnarray}
\hspace*{5pt}\sigma^2_{ij}|\operatorname{rest} &\sim& \operatorname{InvGamma}\Biggl(S/2+c_j,\sum_{s=1}^S
\bigl(y_{ij}(s)-a_j-b_j\mu_{i}(s) \bigr)^2/2+d_j\Biggr),\nonumber\\[-8pt]\\[-8pt]
\tau^2_{j}|\operatorname{rest} &\sim& \operatorname{InvGamma}\bigl(S/2+e,\mathbf{r}_i'Q(\rho_i)\mathbf{r}_i/2+f\bigr),\nonumber
\end{eqnarray}
where $\mathbf{r}_i = \bolds{\mu}_i-W\bolds{\alpha}-\Omega_i\bolds{\beta}$.

The intercept/slope pairs $(a_{j},b_{j})$ have bivariate normal full
conditionals with mean\vspace*{-2pt}
\begin{eqnarray}
V((a_{j},b_{j})'|\operatorname{rest})^{-1} &=& w^{-2}I_2 + \sum_{i=1}^N\Delta_i'\Delta_i/\sigma^2_{ij},\nonumber\\
E((a_{j},b_{j})'|\operatorname{rest}) & = & V((a_{j},b_{j})'|\operatorname{rest})\Biggl(\sum_{i=1}^N\Delta_i'\mathbf{y}_{ij}/\sigma^2_{ij} \Biggr),\nonumber
\end{eqnarray}
where $\Delta_i = (\mathbf{1}, \bolds{\mu}_i)$. The regression coefficients
$\bolds{\alpha}$ and $\bolds{\beta}$ have multivariate normal full
conditionals with\vspace*{-1pt}
\begin{eqnarray}
V(\bolds{\alpha}|\operatorname{rest})^{-1} &=& w^{-2}I_{p_s} +\sum_{i=1}^NW'Q(\rho_i)W/\tau^2_i,\nonumber\\
E(\bolds{\alpha}|\operatorname{rest}) & = & V(\bolds{\beta}_s|\operatorname{rest})\mathbf{X}_s'
\sum_{i=1}^NQ(\rho_i) (\bolds{\mu}_i-\Omega_i'\bolds{\beta})/\tau^2_i\nonumber
\end{eqnarray}
and\vspace*{-1pt}
\begin{eqnarray}
V(\bolds{\beta}|\operatorname{rest})^{-1} &=& w^{-2}I_{p} + \sum_{i=1}^N\Omega_i'Q(\rho_i)\Omega_i/\tau^2_i,\nonumber\\
E(\bolds{\beta}|\operatorname{rest}) & = & V(\bolds{\beta}|\operatorname{rest})
\sum_{i=1}^N\Omega_i'Q(\rho_i) (\bolds{\mu}_i-W\bolds{\alpha})/\tau^2_i.\nonumber
\end{eqnarray}
The remaining parameters $\{c_j\}$, $\{d_j\}$, $e$, $f$ and $g$ are
updated using Metropolis sampling with Gaussian candidate distributions
tuned to give acceptance ratios near 0.40.

\section{Simulation study}\label{s:sim}

In this section we conduct a simulation study to demonstrate the
effects of spatial correlation and informative missingness on the
analysis of patient-level fixed effects. For computational purposes we
assume only one quadrant (i.e., half jaw) for each patient leaving
$S=42$, that there are no spatial covariates $W$, and the same CAR
spatial association parameter for each patient, that is, $\rho_i=\rho$.
We also assume there is only a single continuous response. Data are
generated from the model
%
\begin{eqnarray}
P\bigl(y_i(s) = \mbox{observed}\bigr) &=& 1-\Phi\bigl(a_0 + b_0\mu_i(s) \bigr),\nonumber\\[-8pt]\\[-8pt]
y_i(s)|y_i(s) \mbox{ observed} & \sim& N\bigl(a_1+b_1\mu_i(s),\sigma_i^2\bigr),\nonumber
\end{eqnarray}
where $\bolds{\mu}_i\sim N([x_i'\beta]\mathbf{1}_S$,
$\tau_i^2Q^{-1}(\rho)$).
Each simulated data set contains data generated from this model for
$N=50$ patients. The $p=6$ patient-level covariates $\mathbf{x}_i$ are
generated independently from the standard normal distribution and the
regression coefficients are $\bolds{\beta}= (0,0,0,1,2,3)/20$. Finally,
$a_1=b_1=1$ and $a_0=-1$.

$M=100$ data sets are generated from each of six designs specified by
varying the true value of the covariance parameters $\sigma_i^2$,
$\tau
_i^2$ and $\rho$ and the missing data mechanism $b_0$:
\begin{itemize}
\item\textit{Design 1}: $\rho=0.0$, $b_0=0$ and $\sigma_i^2=\tau_i^2=1$,
\item\textit{Design 2}: $\rho=0.9$, $b_0=0$ and $\sigma_i^2=\tau_i^2=1$,
\item\textit{Design 3}: $\rho=0.9$, $b_0=0$ and $\sigma_i^2=\tau_i^2=$ 1.5*I($i$ is odd)${}+{}$0.5,
\item\textit{Design 4}: $\rho=0.9$, $b_0=1$ and $\sigma_i^2=\tau_i^2=1$,
\item\textit{Design 5}: $\rho=0.9$, $b_0=1$ and $\sigma_i^2=\tau_i^2=$ 1.5*I($i$ is odd)${}+{}$0.5,
\item\textit{Design 6}: $\rho=0.5$, $b_0=1$ and $\sigma_i^2=\tau_i^2=$ 1.5*I($i$ is odd)${}+{}$0.5.
\end{itemize}
Observations within patients are independent under the first design and
spatially correlated under all other designs. The variances are the
same for all patients under Design~2 and vary across patients for
Design~3. Designs~4 and~5 are similar to Designs~2 and~3, except that
the locations of missing observations are informative with $b_0=1$.
Design~6 is the same as Design~5, except with moderate spatial
association $\rho=0.5$.

We analyze each simulated data set using five models:
\begin{itemize}
\item\textit{Model 1}: Linear regression,
${\bar y}_i = \sum_{s\in S_i}y_i(s)/|S_i|\sim N (x_i'\beta,\sigma^2)$,
\item\textit{Model 2}: Section~\ref{s:model}'s spatial model without
informative missingness or patient-specific variances, that is,
$b_0=0$, $\sigma_i^2=\sigma^2$ and $\tau_i^2=\tau^2$,
\item\textit{Model 3}: Section~\ref{s:model}'s spatial model with
patient-specific variances but without informative missingness, that
is, $b_0=0$,
\item\textit{Model 4}: Section~\ref{s:model}'s spatial model with
informative missingness but without patient-specific variances, that
is, $\sigma_i^2=\sigma^2$ and $\tau_i^2=\tau^2$,
\item\textit{Model 5}: Section~\ref{s:model}'s full spatial model,
\end{itemize}
where $S_i$ in Model~1 is the set of locations of observed data for
patient $i$. Model~1 ignores spatial associations and missing teeth,
and simply uses each patient's average observed response in a multiple
regression. Models~2--5 explicitly model all observations individually
and account for spatial associations between nearby observations.

The results are presented in Table~\ref{t:sim}. For each model and each
design, we calculate the proportion of the 95\% posterior intervals for
$b_0$ and the regression coefficients that exclude zero. We also
compute the mean squared error and relative bias,
$\mathrm{MSE}=\frac{1}{pM}\sum_{m=1}^M\sum_{j=1}^p (\hat{\beta}_{j}^{(m)}-\beta_j )^2$\vspace*{1pt} and
$\mathrm{RelBias}_j=\frac{1}{M}\sum_{m=1}^M (\hat{\beta}_{j}^{(m)}-\beta_j)/\beta_j$,
where $\hat{\beta}_{j}^{(m)}$ is the posterior mean of
$\beta_j$ for the $m$th simulated data set and $\beta_j$ is the true
value. Relative bias is only presented for the largest coefficient,
$\beta_6$.

%
\begin{table}
\caption{Simulation study results. Column labels ``$b_0$''--``$\beta_6$'' give
the proportion of 95\% intervals that exclude
zero. The Monte Carlo standard errors (not shown) are between 0.007 and
0.045 for 100*\textup{MSE} and between 0.003 and 0.006 for the bias}\label{t:sim}
\begin{tabular*}{\textwidth}{@{\extracolsep{\fill}}lcccccccccd{2.3}@{}}
\hline
\textbf{Design}
& \textbf{Model}
& $\bolds{b_0}$
& $\bolds{\beta_1}$
&$\bolds{\beta_2}$
&$\bolds{\beta_3}$
&$\bolds{\beta_4}$
&$\bolds{\beta_5}$
&$\bolds{\beta_6}$
& $\bolds{100^{*}\mathrm{MSE}}$
& \multicolumn{1}{c@{}}{$\bolds{\mathrm{RelBias}_6}$}\\
\hline
1 & 1 & -- & 0.06 & 0.05 & 0.03 & 0.29 & 0.85 & 1.00 & 0.103 & 0.022\\
& 2 & -- & 0.05 & 0.07 & 0.04 & 0.34 & 0.86 & 1.00 & 0.103 & 0.021\\
& 3 & -- & 0.05 & 0.06 & 0.04 & 0.35 & 0.86 & 1.00 & 0.104 & 0.022\\
& 4 & 0.10 & 0.05 & 0.05 & 0.03 & 0.35 & 0.85 & 1.00 & 0.104 & -0.001\\
& 5 & 0.12 & 0.06 & 0.05 & 0.04 & 0.35 & 0.83 & 1.00 & 0.106 & 0.006\\
[3pt]
2 & 1 & -- & 0.03 & 0.04 & 0.01 & 0.13 & 0.52 & 0.80 & 0.289 & 0.042\\
& 2 & -- & 0.03 & 0.05 & 0.02 & 0.15 & 0.56 & 0.80 & 0.287 & 0.034\\
& 3 & -- & 0.03 & 0.03 & 0.03 & 0.17 & 0.48 & 0.79 & 0.298 & 0.045\\
& 4 & 0.06 & 0.03 & 0.06 & 0.01 & 0.16 & 0.52 & 0.83 & 0.285 & 0.037\\
& 5 & 0.06 & 0.03 & 0.05 & 0.01 & 0.16 & 0.46 & 0.81 & 0.297 & 0.043\\
[3pt]
3 & 1 & -- & 0.03 & 0.07 & 0.04 & 0.12 & 0.31 & 0.51 & 0.657 & 0.077\\
& 2 & -- & 0.04 & 0.08 & 0.08 & 0.13 & 0.31 & 0.52 & 0.655 & 0.072\\
& 3 & -- & 0.09 & 0.09 & 0.07 & 0.36 & 0.69 & 0.95 & 0.181 & 0.027\\
& 4 & 0.08 & 0.03 & 0.08 & 0.09 & 0.12 & 0.31 & 0.56 & 0.653 & 0.077\\
& 5 & 0.08 & 0.09 & 0.12 & 0.08 & 0.39 & 0.68 & 0.96 & 0.178 & 0.034\\
[3pt]
4 & 1 & -- & 0.04 & 0.08 & 0.05 & 0.14 & 0.43 & 0.70 & 0.266 & -0.150\\
& 2 & -- & 0.06 & 0.05 & 0.07 & 0.16 & 0.43 & 0.76 & 0.267 & -0.146\\
& 3 & -- & 0.04 & 0.06 & 0.06 & 0.18 & 0.42 & 0.72 & 0.265 & -0.141\\
& 4 & 1.00 & 0.04 & 0.10 & 0.04 & 0.18 & 0.58 & 0.89 & 0.278 & 0.048\\
& 5 & 1.00 & 0.02 & 0.06 & 0.05 & 0.19 & 0.58 & 0.86 & 0.267 & 0.026\\
[3pt]
5 & 1 & -- & 0.05 & 0.04 & 0.08 & 0.07 & 0.19 & 0.26 & 0.780 & -0.229\\
& 2 & -- & 0.11 & 0.07 & 0.12 & 0.15 & 0.26 & 0.34 & 0.725 & -0.217\\
& 3 & -- & 0.12 & 0.11 & 0.18 & 0.31 & 0.67 & 0.89 & 0.221 & -0.075\\
& 4 & 1.00 & 0.06 & 0.10 & 0.09 & 0.16 & 0.46 & 0.71 & 0.693 & 0.187\\
& 5 & 1.00 & 0.06 & 0.09 & 0.08 & 0.29 & 0.66 & 0.94 & 0.193 & 0.023\\
[3pt]
6 & 1 & -- & 0.06 & 0.04 & 0.07 & 0.10 & 0.28 & 0.47 & 0.409 & -0.200\\
& 2 & -- & 0.16 & 0.11 & 0.12 & 0.18 & 0.43 & 0.66 & 0.383 & -0.191\\
& 3 & -- & 0.07 & 0.09 & 0.11 & 0.45 & 0.89 & 1.00 & 0.086 & -0.062\\
& 4 & 1.00 & 0.09 & 0.07 & 0.11 & 0.30 & 0.76 & 0.97 & 0.279 & 0.130\\
& 5 & 1.00 & 0.06 & 0.07 & 0.08 & 0.61 & 0.96 & 1.00 & 0.070 & 0.024\\
\hline
\end{tabular*}
\end{table}

Data for the first design are generated without spatial association or
informative missingness. In this case all five models give nearly
identical results, demonstrating that the spatial models are able to
approximate the simple regression model if appropriate. The five models
are also nearly identical for Design~2 where the data are generated
with spatial correlation and the same variances for each patient. In
this case the patient means ${\bar y}_i$ are Gaussian with mean
$a_1+x_i'\beta$ and the same variances, satisfying the usual regression
assumptions.

The linear regression model does not perform well for Design~3's
spatial model with patient-dependent variances. In this case the
patient means ${\bar y}_i$ have different variances, violating the
usual regression assumptions. The spatial models that allow for
patient-dependent variances (Models~3 and~5) give dramatic improvements
in both power and mean squared error compared to the homoskedastic models.

The locations of missing observations are informative for Designs~4, 5
and~6. For these two designs the models (Models~1--3) that do not
account for informative missingness are biased for $\beta_6$. The
models that allow for informative location consistently identify $b_0$
as nonzero (power~1.0 in all three designs), which alleviates the bias
for the nonnull predictors and improves power. Design~5 has both
informative missingness and patient-dependent variances, common traits
of periodontal data. In this case our full model is more than three
times more powerful for $\beta_6$ (0.26 to 0.94) and has roughly one
fourth the mean squared error (0.193 to 0.780) of the usual nonspatial
regression approach.

\section{Analysis of periodontal data}\label{s:results}

%
\begin{table}[b]
\caption{Posterior 95\% intervals for models assuming
variances $\sigma_{ij}^2$ and $\tau_i^2$ are constant across patients.
``Spatial'' models take $\rho\ne0$ and models with informative missing
teeth (``Info missing'') have $b_0 \ne0$}\label{t:coef1}
\begin{tabular*}{\textwidth}{@{\extracolsep{\fill}}lcccc@{}}
\hline
\textbf{Spatial} & \textbf{No} & \textbf{No} & \textbf{Yes} & \textbf{Yes}\\
\textbf{Info missing} & \textbf{No} & \textbf{Yes} & \textbf{No} & \textbf{Yes}\\
\hline
Age & ($-$0.002, 0.022) & (0.009, 0.033) & (0.000, 0.079) & (0.036, 0.115)\\
Female & ($-$0.129, $-$0.104) & ($-$0.128, $-$0.103) & ($-$0.181, $-$0.103) &($-$0.173, $-$0.096)\\
BMI & ($-$0.016, 0.007) & ($-$0.014, 0.011) & ($-$0.048, 0.030) & ($-$0.033, 0.046)\\
Smoker & (0.028, 0.051) & (0.028, 0.051) &(0.014, 0.091) & (0.010, 0.088)\\
Hba1c & (0.114, 0.139) & (0.118, 0.143) & (0.123, 0.199) & (0.128, 0.207)\\
[3pt]
$a_0$: missing & -- & ($-$1.390, $-$1.239) & -- & ($-$1.349, $-$1.172)\\
$a_1$: CAL & (1.008, 1.052) & (1.021, 1.087) & (0.993, 1.112) &(1.002, 1.139)\\
$a_2$: PPD & (1.015, 1.055) & (1.034, 1.101) & (1.092, 1.214) &(1.104, 1.139)\\
$a_3$: BOP & ($-$0.369, $-$0.323) & ($-$0.359, $-$0.312) & ($-$0.399, $-$0.327) &($-$0.394, $-$0.309)\\
$b_0$: missing & -- & (0.432, 0.513) & -- & (0.434, 0.544)\\
$b_2$: PPD & (1.144, 1.178) & (1.131, 1.160) & (1.021, 1.047) &(1.014, 1.043)\\
$b_3$: BOP & (0.473, 0.510) & (0.475, 0.510) & (0.510, 0.547) &(0.434, 0.544)\\
[3pt]
$\rho$ & -- & -- & (0.972, 0.978) & (0.972, 0.978)\\
$\tau$ & (1.454, 1.499) & (1.464, 1.508)& (0.832, 0.870) & (0.838, 0.874)\\
$\sigma_1$: CAL & (0.942, 0.961) & (0.972, 0.978)& (0.881, 0.900) & (0.935, 0.953)\\
$\sigma_2$: PPD & (0.106, 0.182) & (0.177, 0.219)& (0.454, 0.486) & (0.464, 0.493)\\
[3pt]
Tooth 2 & ($-$0.027, 0.038) & ($-$0.038, 0.034) & ($-$0.063, 0.027) &($-$0.072, 0.022)\\
Tooth 3 & (0.037, 0.101) & (0.022, 0.091) & ($-$0.007, 0.106) &($-$0.025, 0.100)\\
Tooth 4 & (0.174, 0.241) & (0.179, 0.253) & (0.106, 0.242) & (0.138, 0.275)\\
Tooth 5 & (0.237, 0.306) & (0.263, 0.339) & (0.214, 0.369) & (0.296, 0.451)\\
Tooth 6 & (0.751, 0.825) & (0.849, 0.932) & (0.597, 0.773) & (0.853, 1.037)\\
Tooth 7 & (0.866, 0.954) & (0.955, 1.048) & (0.597, 0.773) & (0.986, 1.151)\\
Gap & (0.953, 1.001) & (0.938, 0.994) & (0.992, 1.030) & (0.986, 1.022) \\
Maxilla & ($-$0.289, $-$0.246) & ($-$0.293, $-$0.247) & ($-$0.376, $-$0.235) &($-$0.363, $-$0.211)\\
\hline
\end{tabular*}
\end{table}

The motivating data were collected from a clinical study
[Fernandes et al. (\citeyear{FernandezEtAl2009})] conducted by the
Center for Oral Health Research (COHR) at
the Medical University of South Carolina (MUSC). The relationship
between periodontal disease and diabetes level has been previously
studied in the dental literature [Faria-Almeida, Navarro and Bascones
(\citeyear{FariaalmeidaNavarroBascones2006}),
Taylor and Borgnakke (\citeyear{TaylorBorgnakke2008})]. The objective
of this study was to explore the
relationship between periodontal disease and diabetes level (determined
by the popular marker HbA1c, or ``glycosylated hemoglobin'') in the
Type-2 diabetic adult (13 years or older) Gullah-speaking
African-American population residing in the coastal sea-islands of
South Carolina. Since this is part of an ongoing study, we selected 199
patients with complete covariate information and with at least 50\%
responses available.

For each patient CAL, PPD and BOP are measured at six locations on each
nonmissing tooth, as shown in Figure~\ref{f:data}. Additionally,
several patient-level covariates were obtained, including age (in
years), gender (1${}={}$Female, 0${}={}$Male), body mass index or BMI (in kg$/\mbox{m}^{2}$),
smoking status (1${}={}$a smoker, 0${}={}$never) and HbA1c (1${}={}$High,
0${}={}$controlled). We also include spatial covariates for the site in the
gap between teeth (1${}={}$in the gap, 0${}={}$on the side of a tooth), jaw
(1${}={}$maxilla, 0${}={}$mandible) and six tooth number indictors with the first
tooth (front of the mouth, Figure~\ref{f:data}) serving as the
reference tooth. All covariates are standardized to have mean zero and
variance one. The spatial adjacency structure is shown in Figure~\ref
{f:data}; we consider neighboring sites on the same tooth as well as
neighboring sites on the consecutive teeth to be adjacent.

We begin by fitting several models with the same variances for all
patients, that is, $\sigma_{ij}^2=\sigma_j^2$, $\tau_{i}^2=\tau^2$ and
$\rho_{i}^2=\rho^2$. We fit four models by assuming spatial association
($\rho\sim \operatorname{Unif}[0,1]$) and independence ($\rho=0$), and assuming
missing teeth are informative ($b_0\ne0$) and not informative
$(b_0=0)$. Table~\ref{t:coef1} gives posterior 95\% intervals for
several parameters. The slopes $b_j$ for pocket depth and bleeding on
probing (as described in Section~\ref{s:ID}, slope for attachment loss
is fixed at one) are significantly positive for all models, suggesting
strong positive associations between the three responses. Several
covariates are significant for all models, including patient effects
gender, smoking status and HbA1c status, indicators of a site in the
gap between teeth and a site on the upper jaw, and several tooth number
indictors with sites in the back of the mouth having higher mean responses.

\begin{figure}

\includegraphics{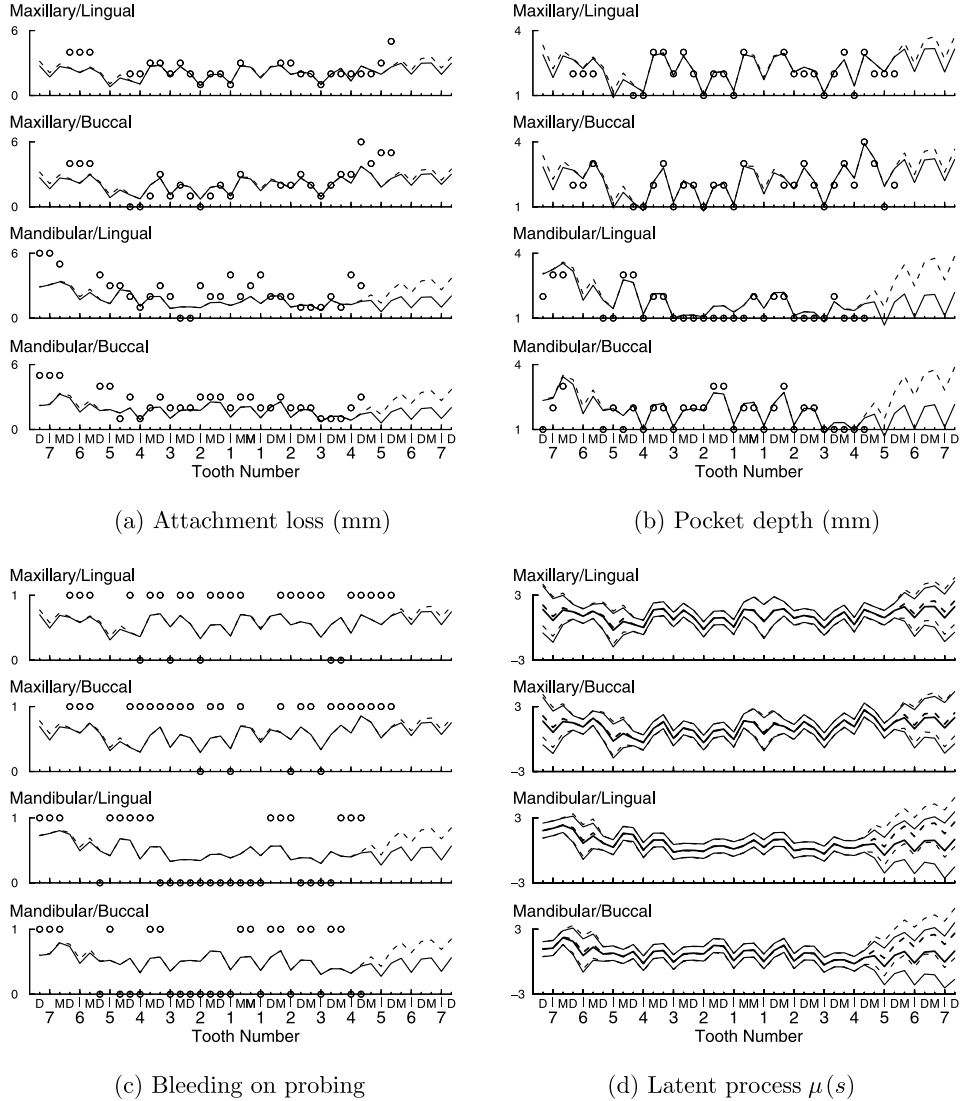}

\caption{Panels \textup{(a)}--\textup{(c)} plot the data (dots) and posterior mean of the
expected response (lines) for a typical patient. Panel \textup{(d)} plots the
posterior mean (bold) and 95\% interval (thin) for the latent spatial
process $\bolds{\mu}(s)$. All plots include results for both the
model with
(dashed line) and without (solid line) informative observation
location. ``Maxillary'' and ``Mandibular'' refer to upper and lower
jaws respectively, while ``buccal'' and ``lingual'' refer to the cheek
and the tongue sides of the teeth, respectively.}\label{f:onesub}
\vspace*{-7pt}
\end{figure}

The slope relating the latent spatial process with the probability of a
missing tooth, $b_0$, is also significantly positive. This matches the
intuition that patients with poor periodontal health generally have
more missing teeth. Figure~\ref{f:onesub} plots the data and fitted
values for a typical patient to illustrate the effects of accounting
for informative missing teeth. This plot compares the spatial models
with $b_0$ set to zero (solid lines) and $b_0$ not set to zero (dashed
lines). The posterior means [Figure~\ref{f:onesub}(a)--(c)]
and credible sets [\ref{f:onesub}(d)] are nearly identical for
observations on nonmissing teeth. However, for missing teeth the fitted
values for all three responses are larger (worse periodontal health)
when accounting for informative observation location.

%

Accounting for informative observation location also affects the
patient effect for age. The 95\% interval ignoring spatial association
and informative observations location is ($-$0.002, 0.022), compared to
(0.036, 0.115) for the full model. The measures of periodontal disease
are cumulative, so it seems reasonable that age should be an important
predictor. Our data show a relationship between age and the number of
missing teeth; patients that are younger than 54 (the mean age) have an
average of 135.8 ($\mathrm{sd}=20.1$) observations and patients that are older
than 54 have an average of 124.7 ($\mathrm{sd}=22.2$) observations. By accounting
for this relationship, we identify age as a significant predictor of
periodontal health.

\begin{figure}

\includegraphics{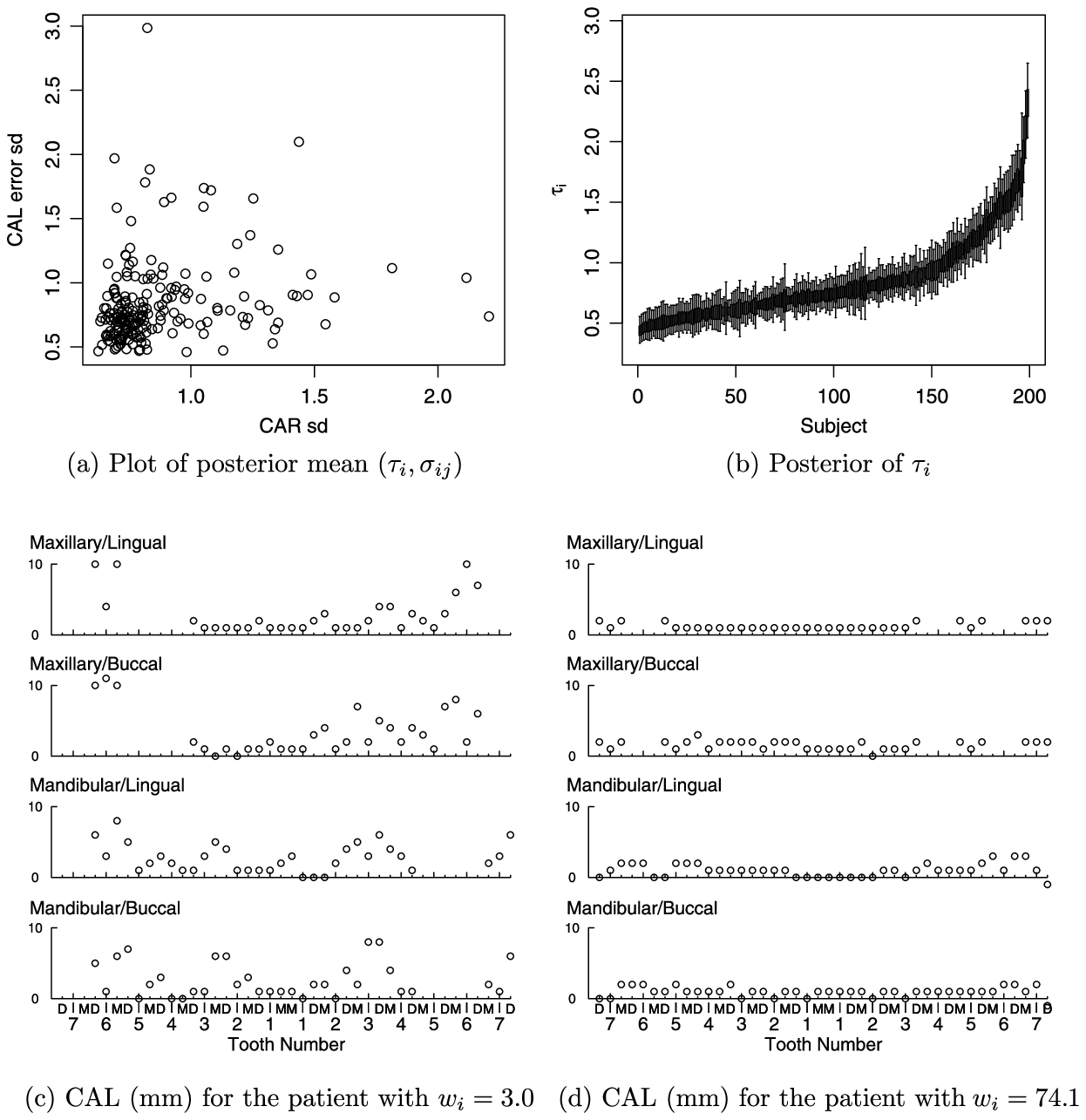}

\caption{Panel \textup{(a)} gives the posterior medians of the patient-specific
standard deviations $(\tau_i,\sigma_{ij})$ for attachment loss, and
panel \textup{(b)} plots the posterior of the CAR standard deviations $\tau_i$
(the horizontal lines in the $i$th column are the posterior 0.025,
0.25, 0.5, 0.75 and 0.975 quantiles for $\tau_i$). Panels \textup{(c)} and \textup{(d)}
plot the attachment loss for the patients with smallest and largest
weights $w_i$, respectively. ``Maxillary'' and ``Mandibular'' refer to
upper and lower jaws, respectively, while ``buccal'' and ``lingual''
refer to the cheek
and the tongue sides of the teeth, respectively.}\label{f:sds}
\end{figure}

Section~\ref{s:sim}'s simulation study shows that the fixed effects can
also be affected if patients have different spatial covariances. To
explore this possibility for our periodontal data, we apply
Section~\ref
{s:model}'s model with variances $\sigma_{ij}^2$ and $\tau_i^2$ varying
across patients. Figure~\ref{f:sds}(a) and~(b) summarize the
posteriors of the variance parameters. Here we see considerable
variation across patients; Figure~\ref{f:sds}(b) shows that the posterior
95\% intervals for $\tau_i$ are nonoverlapping for patients with small
and large~$\tau_i$.

%

Section~\ref{s:diag}'s $w_i$ diagnostic in (\ref{w}) indicates which
patients are the most influential on the regression coefficients. The
$w_i$ (computed using only the CAL error variance) have median 28.0 and
vary greatly across patients with 95\% interval (5.8, 61.7).
Figure~\ref{f:sds}(c) and~(d) plot CAL for the patients with
the smallest and largest~$w_i$. The responses for the patient with
smallest $w_i$ vary considerably from site-to-site within the mouth,
with attachment loss ranging from 0 to 11~mm. Information about the
patient-level covariates accumulate via the mean of the latent
parameters $\bolds{\mu}_i$; due to spatial variability, the mean is quite
uncertain for this patient and, thus, this patient provides little
information about $\bolds{\beta}$.
In contrast, the patient with largest $w_i$
is stable from site-to-site, providing reliable information the mean of
$\bolds{\mu}_i$ and thus about $\bolds{\beta}$.

Table~\ref{t:coef2} gives the 95\% posterior intervals for several
parameters from the model with patient-dependent variances. Comparing
the spatial models with informative missingess, the results for the
patient-level covariates are fairly similar for the models with and
without patient-dependent variances (i.e., the final columns of Ta\-bles~\ref{t:coef1}
and~\ref{t:coef2}). However, we note that the width of the
credible intervals are smaller for the model with patient-dependent variances.

%
\begin{table}
\caption{Posterior 95\% intervals for models assuming
variances $\sigma_{ij}^2$ and $\tau_i^2$ vary across patients.
``Spatial'' models take $\rho\ne0$ and models with informative missing
teeth (``Info missing'') have $b_0 \ne0$}\label{t:coef2}
\begin{tabular*}{\textwidth}{@{\extracolsep{\fill}}lcccc@{}}
\hline
\textbf{Spatial} & \textbf{No} & \textbf{No} & \textbf{Yes} & \textbf{Yes}\\
\textbf{Info missing} & \textbf{No} & \textbf{Yes} & \textbf{No} & \textbf{Yes}\\
\hline
Age & ($-$0.019, 0.007) & ($-$0.013, 0.013) & ($-$0.002, 0.057) & (0.023, 0.086)\\
Female & ($-$0.114, $-$0.086) & ($-$0.115, $-$0.087) & ($-$0.159, $-$0.096) & ($-$0.168, $-$0.104)\\
BMI & ($-$0.008, 0.019) & ($-$0.006, 0.020) & ($-$0.015, 0.040) & ($-$0.009, 0.050)\\
Smoker & (0.038, 0.062) & (0.037, 0.061) &(0.021, 0.078) & (0.019, 0.075)\\
Hba1c & (0.089, 0.114) & (0.090, 0.118) & (0.095, 0.155) & (0.106, 0.171)\\
[3pt]
$a_0$: missing & -- & ($-$1.154, $-$1.004) & -- & ($-$1.201, $-$1.040)\\
$a_1$: CAL & (0.859, 0.921) & (0.851, 0.930) & (0.855, 0.942) &(0.892, 0.989)\\
$a_2$: PPD & (0.899, 0.960) & (0.889, 0.966) & (0.920, 1.012) &(0.958, 1.058)\\
$a_3$: BOP & ($-$0.425, $-$0.373) & ($-$0.424, $-$0.370) & ($-$0.482, $-$0.414) &($-$0.464, $-$0.394)\\
$b_0$: missing & -- & (0.265, 0.378) & -- & (0.294, 0.410)\\
$b_2$: PPD & (1.002, 1.014) & (1.001, 1.014) & (1.017, 1.036) &(1.013, 1.031)\\
$b_3$: BOP & (0.462, 0.499) & (0.463, 0.498) & (0.518, 0.559) &(0.521, 0.560)\\
[3pt]
$\rho$ & -- & -- & (0.954, 0.962) & (0.958, 0.965)\\
Tooth 2 & ($-$0.051, 0.010) & ($-$0.050, 0.019) & ($-$0.054, 0.023) &($-$0.063, 0.016)\\
Tooth 3 & (0.022, 0.081) & (0.021, 0.086) & (0.007, 0.105) &($-$0.009, 0.095)\\
Tooth 4 & (0.183, 0.246) & (0.190, 0.258) & (0.138, 0.251) & (0.141, 0.254)\\
Tooth 5 & (0.307, 0.374) & (0.318, 0.391) & (0.267, 0.385) & (0.291, 0.412)\\
Tooth 6 & (0.776, 0.850) & (0.825, 0.909) & (0.634, 0.763) & (0.753, 0.899)\\
Tooth 7 & (0.854, 0.940) & (0.902, 0.991) & (0.630, 0.769) & (0.782, 0.938)\\
Gap & (0.887, 0.929) & (0.886, 0.932) & (0.894, 0.926) & (0.890, 0.922) \\
Maxilla & ($-$0.280, $-$0.241) & ($-$0.278, $-$0.238) & ($-$0.305, $-$0.208) &($-$0.301, $-$0.201)\\
\hline
\end{tabular*}
\end{table}

Finally, we conducted a sensitivity analysis to determine the effect of
modeling assumptions for the full spatial model with patient-dependent
variances and informative observation location. We modified the
analysis by changing the reference group with slope $b_j$ fixed to one
from CAL to PPD and BOP, changing the hyperparameters $u=v=0.1$ to
$u=v=0.0001$, and changing the hyperparameter $w=10$ to $w=1000$. The
posterior 95\% intervals are given in Table~\ref{t:sense} for the
patient effects, scaled by $b_1$ for comparison across reference group.
The modification with the largest effect is changing the reference
group from CAL to BOP. The patient level effects are generally closer
to zero using BOP as the reference group. Despite this change in scale,
the signs of the coefficients and the subset of coefficients with
intervals that exclude zero remains the same as the original analysis.

%
\begin{table}
\caption{Posterior 95\% intervals for the patient effects
for various modeling/prior choices for the full model with spatial
correlation, patient dependent variance and informative observation
location. ``Ref group'' refers to the response that has slope $b_j$
fixed to one, ``$u, v$'' and ``$w$'' are the hyperparameters for the
covariance parameters and regression coefficients, respectively, as
described in Section~\protect\ref{s:ID}. To facilitate comparison across
reference groups, the intervals for $\bolds{\beta}/b_1$ are
presented}\label{t:sense}
\begin{tabular*}{\textwidth}{@{\extracolsep{\fill}}lcccc@{}}
\hline
\textbf{Ref group} & \textbf{CAL} & \textbf{PPD} & \textbf{BOP} & \textbf{CAL} \\
$\bolds{u, v}$ & $\bolds{0.1}$ & $\bolds{0.1}$ & $\bolds{0.1}$ & $\bolds{0.001}$ \\
$\bolds{w}$ & $\bolds{10}$ & $\bolds{10}$ & $\bolds{10}$ & $\bolds{10}$ \\
\textbf{Spatial grid} & $\bolds{1}$ & $\bolds{1}$ & $\bolds{1}$ & $\bolds{1}$\\
\hline
Age & (0.023, 0.086) & (0.025, 0.088) & (0.007, 0.079) & (0.039, 0.110) \\
Female & ($-$0.168, $-$0.104) & ($-$0.175, $-$0.108) & ($-$0.052, $-$0.031) & ($-$0.169, $-$0.097) \\
BMI & ($-$0.009, 0.050) & ($-$0.012, 0.052) & ($-$0.003, 0.015) & ($-$0.015, 0.054) \\
Smoker & (0.019, 0.075) & (0.020, 0.077) &(0.005, 0.023) & (0.009, 0.075) \\
Hba1c & (0.106, 0.171) & (0.110, 0.174) & (0.033, 0.054) & (0.122, 0.190) \\
[12pt]
\textbf{Ref group} & \textbf{CAL} & \textbf{CAL} & \textbf{CAL} & \\
$\bolds{u, v}$ & $\bolds{0.1}$ & $\bolds{0.1}$ & $\bolds{0.1}$ &\\
$\bolds{w}$ & $\bolds{1000}$ & $\bolds{10}$ & $\bolds{10}$ & \\
\textbf{Spatial grid} & $\bolds{1}$ & $\bolds{2}$ & $\bolds{3}$ & \\
\hline
Age     & (0.039, 0.109) & (0.023, 0.068) & (0.038, 0.105)&\\
Female  & ($-$0.174, $-$0.096)& ($-$0.150, $-$0.103)& ($-$0.167, $-$0.097)&\\
BMI     & ($-$0.017, 0.050)& ($-$0.009, 0.035)& ($-$0.011, 0.054)&\\
Smoker  & (0.009, 0.076)& (0.024, 0.067)& (0.010, 0.074)&\\
Hba1c   & (0.119, 0.191)& (0.117, 0.164)& (0.111, 0.176)&\\
\hline
\end{tabular*}
\end{table}

Also, we consider modifying the adjacency structure shown by the gray
lines in Figure~\ref{f:data} (``spatial grid 1'') in two ways: first by
not considering sites on the opposite side of a tooth to be neighbors
(``spatial grid 2'') to give independent $\operatorname{AR}(1)$ models to the sites on
the buccal and lingual sides of each jaw, and second by considering all
pairs of observations on the same tooth to be neighbors (``spatial grid
3''). To determine how well each of these spatial grids fit our data,
we use the deviance information criteria (DIC) of Spiegelhalter et al.
(\citeyear{SpiegelhalterEtAl2002}).
To compare spatial structures using DIC, we analyze only a
single continuous response, CAL, and do not consider informative
missing teeth. DIC prefers grid 1 (DIC${}={}$66,128) over grids 2 (DIC${}={}$68,168) and 3 (DIC${}={}$68,562).
Table~\ref{t:sense} gives the posterior of
the subject-level effects for the full data analysis using the three
spatial grids; the results are not sensitive to the choice of spatial structure.

\section{Discussion}\label{s:disc}

In this paper we develop a latent factor model for multivariate spatial
periodontal data with a mix of binary and continuous responses. Our
model allows for a different spatial covariance for each patient and
for informative missing teeth. We show using simulated and real data
that accounting for these factors leads to a substantial improvement
for estimating covariate effects compared to standard regression techniques.

We have assumed throughout that the patient's periodontal health can be
captured by a single latent factor. It would be straightforward,
conceptually if not computationally, to include more latent factors.
However, this leads to the problem of selecting the appropriate number
of latent factors, interpreting the roles of the different latent
factors, and understanding the effects of the covariates on the
different latent factors. For these data with three strongly-correlated
responses we prefer the single factor model for computational
simplicity and interpretability. If multiple factors are allowed, the
number of factors could be chosen using the deviance information
criteria. Another approach would be to allow the number of factors to
be unknown. Lopes, Salazar and Gamerman (\citeyear
{LopesSalazarGamerman2008}) and
Salazar, Gamerman and Lopes (\citeyear{SalazarGamermanLopes2009}) use
reversible jump MCMC to account for uncertainty in
the number of latent factors. Extending this approach to our setting
may be complicated by the large number of subjects, since the proposal
density would have to propose spatial models that simultaneously fit
well for all 199 subjects. Another possibility would be to extend the
parameter expansion method of Ghosh and Dunson (\citeyear
{GhoshDunson2008}) to the spatial setting.

We have also assumed that the latent spatial process is Gaussian. For
nonspatial data several authors have proposed methods that avoid
assuming the shared random effects are Gaussian [Lin et al.
(\citeyear{LinEtAl2000}),
Song, Davidian and Tsiatis (\citeyear{SongDavidianTsiatis2002}),
Beunckens et al. (\citeyear{BeunckensEtAl2008}),
Tsonaka, Verbeke and Lesaffre (\citeyear{TsonakaVerbekeLesaffre2009})].
These approaches could be extended to the periodontal setting by
replacing the Gaussian spatial model with a non-Gaussian spatial model
[e.g., Gelfand, Kottas and MacEachern (\citeyear{GelfandKottasMacEachern2005}),
Griffin and Steel (\citeyear{GriffinSteel2006}),
Reich and Fuentes (\citeyear{ReichFuentes2007})].

An area of future work is to apply this method to longitudinal
periodontal data. Periodontal data is often collected repeatedly for a
single patient over time to monitor disease progression.
Reich and Hodges (\citeyear{ReichHodges2008}) propose a spatiotemporal
model for attachment loss. It
should be possible to extend this model to accommodate mixed
multivariate responses and informative missing teeth.

\section*{Acknowledgments}
The authors thank the Center for Oral Health Research (COHR) at the
Medical University
of South Carolina for providing the data and context for this work, in
particular,
Drs. S. London, J. Fernandes, C. Salinas, W.~Zhao, Ms. L. Summerlin
and Ms. P. Hudson.
We also wish to acknowledge several helpful discussions with
Dr.~James Hodges of the University of Minnesota.

\begin{supplement}[id=suppA]
\stitle{Computer code (spatial factor.R)}
\slink[doi]{10.1214/09-AOAS278SUPP}
\slink[url]{http://lib.stat.cmu.edu/aoas/278/spatial\%20factor.R}
\sdescription{In the supplemental file, we include R code to analyze
a single continuous response with informative missingness.
Use of the code is described in the file and is illustrated with an
analysis of a simulated data set.}
\sdatatype{.R}
\end{supplement}

\printaddresses

\end{document}